\documentclass{aa}  

\usepackage{bm}
\usepackage{color}
\usepackage{float}
\usepackage{graphicx,amsmath}
\usepackage{txfonts}
\usepackage{natbib}
\def\wig#1{\mathrel{\hbox{\hbox to 0pt{%
          \lower.6ex\hbox{$\sim$}\hss}\raise.4ex\hbox{$#1$}}}}
\def\msol{{\rm M_\odot}}

\def\lsol{{\rm L_\odot}}
\def\mearth{{\rm M_\oplus}}

\def\v{{\rm v}}  % This is to distinguish v and nu
\def\au{{\rm au}}
\def\hg{h_{\rm g}}
\def\hhg{\hat{h}_{\rm g}}
\def\st{\tau_{\rm s}}
\def\sigp{\Sigma_{\rm p}}
\def\sigg{\Sigma_{\rm g}}
\def\hp{h_{\rm p}}
\def\hhp{\hat{h}_{\rm p}}
\def\vk{{\rm v_{\rm K}}}
\def\RH{R_{\rm H}}

\usepackage[normalem]{ulem}

\begin{document}

   \title{The radial dependence of pebble accretion rates: A source of diversity in planetary systems I. Analytical formulation}

   \author{S. Ida
          \inst{1}
          \and
          T. Guillot
          \inst{2}
          \and
          A. Morbidelli
          \inst{2}
          }

   \institute{Earth-Life Science Institute, Tokyo Institute of Technology, Meguro-ku, Tokyo 152-8550, Japan\\
              \email{ida@elsi.jp}
         \and
             Laboratoire J.-L.\ Lagrange, Universit\'e C\^ote d'Azur, Observatoire de la
  C\^ote d'Azur, CNRS, F-06304 Nice, France\\
           }

   \date{DRAFT:  \today}

% \abstract{}{}{}{}{} 
% 5 {} token are mandatory
 
  \abstract
{The classical planetesimal accretion scenario for the formation of planets has recently evolved with the idea that pebbles, centimeter- to meter-sized icy grains migrating in protoplanetary disks, can control planetesimal and/or planetary growth.}
   {We investigate how pebble accretion depends on disk properties and affects the formation of planetary systems}
   {We construct analytical models of pebble accretion onto planetary embryos that consistently account for the mass and orbital evolution of the pebble flow and reflect disk structure.}
   {We derive simple formulas for pebble accretion rates
   in the so-called settling regime for planetary embryos that are more than 100 km in size.
   For relatively smaller embryos or in outer disk regions, the accretion mode is three-dimensional (3D), 
   meaning that the thickness of the pebble flow must be taken into account, and resulting in an accretion rate that is
   independent of the embryo mass. For larger embryos or in inner regions,
   the accretion is in a two-dimensional (2D) mode, i.e., the pebble disk may be considered infinitely thin.
   We show that the radial dependence of the pebble accretion rate is different (even the sign 
   of the power-law exponent changes) for different disk conditions
   such as the disk heating source (viscous heating or stellar irradiation),
   drag law (Stokes or Epstein, and weak or strong coupling), and in the 2D or 3D accretion modes.
   We also discuss the effect of the sublimation and destruction of icy pebbles inside the snow line. }
   {Pebble accretion easily produces a large diversity of planetary systems.
     In other words, to infer the results of planet formation through pebble accretion correctly,
     detailed prescriptions of disk evolution and pebble growth, sublimation, destruction and migration
     are required.}

     
     \titlerunning{Radial dependence of pebble accretion}

   \maketitle
%
%________________________________________________________________

\section{Introduction}

%\textcolor{red}{The introduction has of course to be enlarged !}

The conventional model of planet formation \citep[e.g.,][]{Safronov72,Hayashi85} assumed that the building blocks of planetary bodies are km-sized planetesimals.
However, agglomerating dust grains in a protoplanetary disk 
to form these planetesimals leads to a serious problem, the so-called radial drift barrier (a.k.a. meter-size barrier). 
While small enough grains are strongly coupled to the disk gas,
larger bodies migrate more rapidly through angular momentum loss by aerodynamical gas drag
until they reach kilometric sizes.
The migration of meter-sized bodies is as fast as $\sim 10^{-2}$ \,\au/yr
 \citep[e.g.,][]{Weiden80,Nakagawa81}. 
 
For small dust grains, growth via pairwise collisions is faster than migration
(if collisions result in coagulation rather than rebounding or fragmentation; see section 3.4), 
so that they actually grow {\it in situ}.
When they grow to $O(10)$ cm-sized bodies called pebbles,
migration dominates over growth and they actually start their migration.
Although the migration of pebbles is slower than that of meter-sized bodies, it is
still $\ga 10^{-4}$ \,\au/yr and the migration timescale is shorter by a few orders than
gas disk lifetimes that are observationally inferred.
These bodies cannot bypass the radial drift barrier
through more rapid growth, unless the fluffy structure of icy dust grains
\citep{Okuzumi12} is considered or local disk structure such as long-lived turbulent anticlonic eddies 
\citep[e.g.,][]{Barge95, Chavanis00, Johansen04, Inaba06}
or radial pressure bumps \citep[e.g.,][ and references therein]{JohansenPP6} exists.

However, a swarm of these fast migrating pebbles may cause a traffic jam 
resulting in a streaming instability and form large clumps
that may become 100-1000 km-sized bodies \citep{Youdin05, Johansen07, Johansen12, Johansen15}.
Some of these simulations \citep{Johansen12} suggest that only a few tens of percent of solid
materials could be incorporated into these clumps. 
The clumps would subsequently grow by accreting the migrating pebbles
\citep{LJ12,LJ14a,LJ14b}, a process commonly called pebble accretion.
This idea was applied among others, to the formation of Jovian cores \citep{Levison15},
close-in super-Earths in exoplanetary systems \citep{Chatterjee14,Chatterjee15, Moriarty15},
an explanation of the dichotomy of the solar system \citep{Morbidelli15a},  
and to account for water delivery to the Earth \citep{Morbidelli15b,Sato15}.

Planet formation through planetesimal accretion is a local process, that is,
planetesimals are accreted in a local feeding zone, 
until planets acquire lunar to Martian mass for which type I migration becomes effective 
\citep[e.g.,][]{Tanaka02,Paardekooper11}.
Because Kepler frequency and planetesimal spatial density are higher
in inner regions, the planetesimal accretion timescale is shorter for smaller
orbital radius with a relatively strong dependence ($t_{\rm acc}\propto r^{2-3}$).

On the other hand, in the case of pebble accretion, planets accrete pebbles
migrating from outer regions with a small capture probability \citep{Guillot14},
so that the accreting planets share the common pebble flux as building blocks.
It is thus expected that the $r$-dependence of $t_{\rm acc}$ should be very weak.

Fundamental formulas for pebble accretion rate have been studied in detail \citep{Ormel_Klahr10, Ormel_Kobayashi12, 
LJ12, LJ14b,Guillot14}.
Using these expressions, we investigate the radial dependence of pebble accretion 
rates onto planets, consistently taking pebble growth and migration 
and disk properties into account.
In section 2, we set up a simple empirical disk model
based on detailed
analytical calculations and radiative transfer simulations
that includes both viscous heating and stellar irradiation.
In section 3, we model the collisional growth and radial migration of dust and pebbles.    
Based on the dust and pebble evolution model with the disk model, we derive 
simple analytical formulas of the pebble accretion rate onto planetary embryos.
In section 4, we provide expressions for the pebble accretion rate as a function of
gas disk accretion rate ($\dot{M}_*$) and orbital radius ($r$)
and discuss the relations between the radial dependence and disk properties. 
Section 5 is a summary. 
Lastly, the symbols used in this work are listed in Table~\ref{tab:symbols} in the Appendix.

\section{Protoplanetary disk model}

For simplification, we consider steady accretion disks with a constant $\alpha$
(the $\alpha$-viscosity parameter) and parameterize the disk midplane temperature $T$
and gas surface density $\sigg$ with
power-law functions of the orbital distance $r$ as
\begin{equation}
T \propto r^{-\gamma}
\end{equation}
and
\begin{equation}
\sigg \propto r^{-\xi}.
\end{equation}
Because the radial gradient is important in our arguments, 
the power indexes do not
need to be the same values throughout the entire disk.
The scaling laws that we provide below are the same as those derived by \citet{Chambers09}.
For numerical factors, we use the results from \citet{Garaud07} and \citet{Oka11}.

As we show here, the disk aspect ratio $\hg/r$ (where $\hg$ is
the vertical gas disk scale height) is an important factor for pebble accretion.
We set 
\begin{equation}
\hhg \equiv \frac{\hg}{r} \propto r^{q}.
\label{eq:h_g0}
\end{equation}
We choose to define the scale height $\hg$ so that the vertical gas density 
is $\rho_g(z) \propto \exp(- z^2/2\hg^2)$, or equivalently, 
$\hg = c_s/\Omega$, where $c_s$ is the sound velocity and $\Omega$ is Keplerian frequency ($\Omega = \sqrt{GM_*/r^3}$; $M_*$ is the host star mass), both estimated at the midplane and for a given orbital distance\footnote{In some works \citep[e.g.,][]{Hayashi81,HuesoGuillot2005,Guillot14}, 
$\hg$ is defined by $\rho_g(z) \propto \exp(- z^2/\hg^2)$, 
or equivalently, $\hg = \sqrt{2}c_s/\Omega$. In both cases, these expressions implicitly assume a disk that is vertically isothermal and in hydrostatic equilibrium \citep[see][]{HuesoGuillot2005}.}.

Since $T \propto c_s^2 = \hg^2 \Omega^2$,
\begin{equation}
\gamma = -2q+1.
\end{equation}
From the assumption of steady disk accretion,
$\dot{M}_*=3\pi \sigg \nu = 3\pi \alpha \sigg \hg^2 \Omega$,
is independent of $r$, so that
\begin{equation}
\xi = 2q + 1/2 = - \gamma + 3/2.
\label{eq:xi_q}
\end{equation}
The assumption of a steady accretion is generally good in the inner regions, i.e., 
when $r \ll r_{\rm out}$ where $r_{\rm out}$ is the disk outer edge radius.
In order to simplify the quantitative estimates, we introduce normalized parameters
for the stellar mass $M_*$, stellar luminosity $L_*$,
viscous alpha parameter $\alpha$, disk accretion rate $\dot{M}_*$,
and pebble mass accretion flux through the disk $\dot{M}_{\rm F}$ as
\begin{equation}
\begin{array}{l}
{\displaystyle M_{*0} \equiv \frac{M_*}{1\,\msol},}\; \;
{\displaystyle L_{*0} \equiv \frac{L_*}{1\lsol},}\; \; 
{\displaystyle \alpha_3 \equiv \frac{\alpha}{10^{-3}}, } \; \; \\
{\displaystyle \dot{M}_{*8} \equiv \frac{\dot{M}_*}{10^{-8}\,\msol/{\rm yr}},} \; \;
{\displaystyle \dot{M}_{\rm F4} \equiv \frac{\dot{M}_{\rm F}}{10^{-4}\,\mearth/{\rm yr}}.}
\end{array}
\end{equation}
The scaling factor $\dot{M}_* = 10^{-8}\,\msol/{\rm yr}$ is typical
of classical T~Tauri stars, and 
$\dot{M}_{\rm F4} = 10^{-4}\,\mearth/{\rm yr}$ corresponds to a value that is often estimated from theoretical works (section 3.4).

The disk temperature parameter $\gamma$ (or equivalently $q$) is mostly determined by
the heating source. As shown by, for example, \citet{HuesoGuillot2005} and \citet{Oka11}, viscous heating dominates in the inner disk regions while irradiation from the central star dominates the thermal structure of the outer regions of the disk. 
The disk midplane temperature can be approximated by
$T = \max(T_{\rm vis},T_{\rm irr})$, where $T_{\rm vis}$
and $T_{\rm irr}$ are temperatures determined by viscous heating
and stellar irradiation, respectively.
The detailed results of \citet{Garaud07} and \citet{Oka11} are empirically fitted by
\begin{equation}
T_{\rm vis} \simeq 200 M_{*0}^{3/10} \alpha_3^{-1/5} 
\dot{M}_{*8}^{2/5}
\left(\frac{r}{1\,\au}\right)^{-9/10}\; {\rm K},
\label{eq:T_vis}
\end{equation}
and 
\begin{equation}
T_{\rm irr} \simeq 150 
L_{*0}^{2/7}M_{*0}^{-1/7} \left(\frac{r}{1\,\au}\right)^{-3/7}\; {\rm K},
\label{eq:T_irr}
\end{equation}
where the power exponents are derived by analytical arguments.
In the pre-main sequence stellar evolution phase, when protoplanetary disks are present, $L_* \propto M_*^{1}$ to $M_*^{3}$, implying that $T_{\rm irr}$ increases with $M_*$.
Since it is observationally suggested that mean values of $\dot{M}_*$ is proportional to 
$M_*^2$, $T_{\rm vis}$ would also increase with $M_*$.

In order to derive Eq.~(\ref{eq:T_irr}), we implicitly assumed that 
the disk is vertically optically thin, but radially optically thick.
When the disk is so depleted that the disk becomes optically thin
even in the radial direction, $T \simeq 280 L_{*0}^{1/4} (r/1\,\au)^{-1/2}{\,\rm K}$ 
\citep[e.g.,][]{Hayashi81}.
However, the radially thin condition is realized only for $\dot{M}_* \la 10^{-10}\,\msol/{\rm yr}$
\citep[e.g.,][]{Oka11}.  This corresponds to a very low accretion
rate and accordingly a very low disk gas surface density. 
We hence do not consider the optically thin limit in this paper.   
In the irradiated, radially thick limit, the midplane temperature of
the disk is significantly lower than in the optically thin limit.
When viscous heating becomes weak enough ($\dot{M}_* \la 10^{-8}\,\msol/{\rm yr}$),
the snow line is inside 1\,\au \;\citep[e.g.,][]{Oka11}
\footnote{The reason terrestrial planets in our solar system are dry (almost ice-free)
even though they would have formed in such low-temperature disks
is discussed by \citet{Morbidelli15b} and \citet{Sato15}.}.

The corresponding disk scale heights in both temperature regimes are
\begin{equation}
\hat{h}_{\rm g,vis} = \frac{{h}_{\rm g,vis}}{r}
\simeq 0.027 M_{*0}^{-7/20}
\alpha_3^{-1/10}
\dot{M}_{*8}^{1/5}
\left(\frac{r}{1\,\au}\right)^{1/20},
\label{eq:h_vis}
\end{equation}
and 
\begin{equation}
\hat{h}_{\rm g,irr} = \frac{{h}_{\rm g,irr}}{r} \simeq 
0.024 L_{*0}^{1/7}M_{*0}^{-4/7} \left(\frac{r}{1\,\au}\right)^{2/7}.
\label{eq:h_irr}
\end{equation}
The actual scale height is given by $\hg = \max(h_{\rm g,vis},h_{\rm
  g,irr})$. 
The exponent is thus $q = 1/20$ in viscous regime and $q = 2/7$ in irradiation regime.
In the optically thin limit, 
$\hg$ is given by 
$\hg/r \simeq 0.033(r/1\,\au)^{1/4}(M_*/\msol)^{-1/2}(L_*/\lsol)^{1/8}$.

The assumption of a steady accretion disk enables us to calculate the
gas surface density explicitly as
\begin{equation} 
\begin{array}{ll}
\sigg & 
{\displaystyle = \frac{\dot{M}_* }{3\pi \alpha \hg^2 \Omega}
     = \frac{\dot{M}_*  T_K}{6\pi^2 \alpha \hhg^2 r^2}}\\
 & {\displaystyle = \frac{10^{-5}}{6\pi^2}  
   \dot{M}_{*8}  M_{*0}^{-1/2} \left(\frac{r}{1\,\au}\right)^{1/2}
   \alpha_3^{-1} \hhg^{-2} \; [\msol/\au^2]} \\
 & {\displaystyle = 1.5
   \dot{M}_{*8}  M_{*0}^{-1/2} \left(\frac{r}{1\,\au}\right)^{1/2}
   \alpha_3^{-1} \hhg^{-2} \; [{\rm g/cm}^2]}
   \end{array}
\end{equation}
Substituting Eqs.~(\ref{eq:h_vis}) and (\ref{eq:h_irr}) into the above equation, 
the surface
density in the viscous and irradiation regimes, respectively, becomes 
\begin{equation}
\Sigma_{\rm g,vis} \simeq 
2.1 \times 10^3 M_{*0}^{1/5} \alpha_3^{-4/5}
\dot{M}_{*8}^{3/5}
\left(\frac{r}{1\,\au}\right)^{-3/5}{\rm g\,cm}^{-2},
\label{eq:Sigma_vis}
\end{equation}
and
\begin{equation}
\Sigma_{\rm g,irr} \simeq 2.7 \times 10^3 
L_{*0}^{-2/7}M_{*0}^{9/14}
\alpha_3^{-1}
\dot{M}_{*8}
\left(\frac{r}{1\,\au}\right)^{-15/14}{\rm g\,cm}^{-2}.
\label{eq:Sigma_irr}
\end{equation}
At any point, the surface density can be calculated by 
$\sigg = \min(\Sigma_{\rm g,vis},\Sigma_{\rm g,irr})$.

The boundary between the viscous and irradiation regimes given by $T_{\rm vis} = T_{\rm irr}$ 
corresponds to an orbital distance, 
 \begin{equation}
r_{\rm vis-irr} \simeq 1.8 L_{*0}^{-20/33}M_{*0}^{31/33}
\alpha_3^{-14/33}
\dot{M}_{*8}^{28/33} \au.
\label{eq:r_vis_irr}
\end{equation}
Viscous heating dominates for $r < r_{\rm vis-irr}$ and conversely,
stellar irradiation dominates at larger orbital distances. 
For classical T~Tauri stars, $\dot{M}_*\sim 10^{-8}\,\msol/{\rm yr}$,
implying a value of $r_{\rm vis-irr}\sim 2\,\au$, i.e., in the middle of
the expected planet formation region.
As we show later, the properties of pebble accretion
change significantly between the viscous and irradiation regimes.

Because the water inside pebbles should vaporize inside the so-called 
snow line (defined by the region at which $T \sim 170$ K), the
size of pebbles and properties of pebble accretion should change
when $r<r_{\rm snow}$.  The location of the snow line can be obtained
by 
$r_{\rm snow} \sim \max(r_{\rm snow,vis}, r_{\rm snow,irr})$, where
\begin{equation}
r_{\rm snow,vis} \simeq 1.2 M_{*0}^{1/3}
\alpha_3^{-2/9}
\dot{M}_{*8}^{4/9}\au,
\label{eq:r_snow_vis}
\end{equation}
\begin{equation}
r_{\rm snow,irr} \simeq 0.75 L_{*0}^{2/3}M_{*0}^{-1/3}\au.
\label{eq:r_snow_irr}
\end{equation}
Because $\dot{M}_*$ decreases with time, the snow line
migrates inward as long as it is in the viscous regime.
When the snow line is in the irradiation region, shading
effects may complicate its evolution \citep{Bitsch15a}. For
simplicity, we do not consider this possibility.

In summary, in the viscously heated inner region, 
$\gamma \simeq 9/10,  \xi \simeq 3/5$ and $q \simeq 1/20$,
while in the irradiation outer region, 
$\gamma \simeq 3/7, \xi \simeq 15/14$ and $q \simeq  2/7$.
The transition occurs at a few \au.
Pebble size can also change at a similar location because of ice sublimation.
The size of pebbles is also bound to change in this region owing to ice sublimation.
We see that the mode of accretion of pebbles and gas drag law
also change in the same region, thus making planet formation through
the accretion of pebbles particularly complex.

\section{Pebble accretion rate}
\subsection{Stokes number}

Another important parameter for pebble accretion is the Stokes number
$\st$ which expresses how the motion of pebbles are coupled to that of the circumstellar disk gas
in sub-Keplerian rotation. It is defined by 
\begin{equation}
\st = t_{\rm stop}\Omega,
\end{equation}
where $t_{\rm stop}$ is the stopping time due to gas drag. A general
expression of the stopping time is provided by \citet{Guillot14}, but
in our case, we can consider two limits that depend on the size of
the pebbles considered:
\begin{equation}
t_{\rm stop} = 
\left\{
\begin{array}{ll}
{\displaystyle \frac{\rho_{\rm s} R}{c_s \rho_g} 
= \frac{\rho_{\rm s} R}{\hg \Omega \frac{\sigg}{\sqrt{2\pi} \hg}}
= \frac{\sqrt{2\pi} \rho_{\rm s} R}{\Omega \sigg}}
&  [R \la \frac{9}{4} \lambda_{\rm mfp} {\rm : Epstein}], \\
{\displaystyle \frac{4\rho_{\rm s} R^2}{9c_s \lambda_{\rm mfp} \rho_g}
} & [R \ga \frac{9}{4} \lambda_{\rm mfp} {\rm : Stokes}], 
\end{array}
\right. 
\label{eq:t_stop}
\end{equation}
where $\lambda_{\rm mfp}$ is the mean free path, $\rho_{\rm s}$ and
$R$ are the bulk density and physical radius of a dust particle, respectively; 
we used the spatial gas density at the midplane of the disk,
$\rho_g \simeq \sigg/\sqrt{2 \pi} \hg$. 

According to Eq.~(\ref{eq:t_stop}), the Stokes number is given by
\begin{equation}
\st = 
\left\{
\begin{array}{ll}
{\displaystyle  \frac{\sqrt{2\pi} \rho_{\rm s} R}{\sigg} }
& [{\rm Epstein}] ,\\
{\displaystyle \frac{4 \rho_{\rm s} R^2}{9 \rho_g \hg \lambda_{\rm mfp} }
= \frac{4 \rho_{\rm s} \sigma R^2}{9 \mu m_H  \hg}
 }
& [{\rm Stokes}].
\end{array}
\right. 
\label{eq:tau_s}
\end{equation}
where $\sigma$ ($\simeq 2 \times 10^{-15}{\rm cm}^2$) is the collisional cross section
for $H_2$, 
$m_H$ ($\simeq 1.67 \times 10^{-24}{\rm g}$) is the mass of hydrogen, 
$\mu$ ($\simeq 2.34$) is the mean molecular weight.
In the viscous regime with $\Sigma_{\rm g,vis}$ (Eq.~(\ref{eq:Sigma_vis})) and 
$h_{\rm g,vis}$ (Eq.~(\ref{eq:h_vis})), the Stokes number is explicitly given by
\begin{equation}
\st = 
\left\{
\begin{array}{l}
\frac{\sqrt{2\pi} \rho_{\rm s} R}{\sigg} \sim
 1.3 \times 10^{-3} M_{*0}^{-1/5} \alpha_3^{4/5} \dot{M}_{*8}^{-3/5} \rho_{s1} 
  \left(\frac{r}{1\,\au}\right)^{3/5}
\left(\frac{R}{1{\, \rm cm}}\right) \\
 \hspace*{5cm} [{\rm Epstein}], \\
\frac{4 \rho_{\rm s} \sigma R^2}{9 \mu m_H  \hg} \sim
 5.6  \times 10^{-4}
M_{*0}^{7/20}\alpha_3^{1/10}\dot{M}_{*8}^{-1/5} 
\rho_{s1} \left(\frac{r}{1\,\au}\right)^{-21/20}
\left(\frac{R}{1{\, \rm cm}}\right)^2 \\
  \hspace*{5cm} [{\rm Stokes}],
\end{array}
\right. 
\label{eq:tau_s2}
\end{equation}
where $\rho_{s1} = \rho_s/1{\rm g\,cm}^{-3}$.
In the irradiation regime with
$\Sigma_{\rm g,irr}$ (Eq.~(\ref{eq:Sigma_irr})) and 
$h_{\rm g,irr}$ (Eq.~(\ref{eq:Sigma_irr})),
\begin{equation}
\st = 
\left\{
\begin{array}{l}
\frac{\sqrt{2\pi} \rho_{\rm s} R}{\sigg} \sim
 0.93 \times 10^{-3} L_{*0}^{2/7}M_{*0}^{-9/14} \alpha_3 \dot{M}_{*8}^{-1} \rho_{s1} 
  \left(\frac{r}{1\,\au}\right)^{15/14}
\left(\frac{R}{1{\rm cm}}\right) \\
  \hspace*{5cm} [ {\rm Epstein}], \\
\frac{4 \rho_{\rm s} \sigma R^2}{9 \mu m_H  \hg} \sim
 6.3 \times 10^{-4}
L_{*0}^{-1/7}M_{*0}^{4/7}
\rho_{s1} \left(\frac{r}{1\,\au}\right)^{-9/7}
\left(\frac{R}{1{\rm cm}}\right)^2 \\
  \hspace*{5cm} [{\rm Stokes}].
\end{array}
\right. 
\label{eq:tau_s3}
\end{equation}
The mean free path is given by
\begin{equation}
\begin{array}{ll}
\lambda_{\rm mfp}  & {\displaystyle \simeq \frac{\mu m_H}{\sigma \rho_g} 
\simeq \frac{\sqrt{2\pi} \mu m_H \hg}{\sigma \sigg}
 } \\
     & \sim
\left\{
\begin{array}{l}
 1.0  M_{*0}^{-11/20} 
 \alpha_3^{7/10}
\dot{M}_{*8}^{-2/5}
\left(\frac{r}{1\,\au}\right)^{33/20}
{\rm cm} \\
\hspace*{3.5cm} [\mbox{viscous region}], \\
 0.65 L_{*0}^{3/7}M_{*0}^{-17/14} 
 \alpha_3
\dot{M}_{*8}^{-1}
\left(\frac{r}{1\,\au}\right)^{33/14}{\rm cm} \\
\hspace*{3.5cm} [\mbox{irradiation region}].\\
\end{array}
\right.
\end{array}
\label{eq:l_g}
\end{equation}
Since the $r$-dependence of $\lambda_{\rm mfp}$ is relatively strong,
a dust grain in the Epstein regime migrating inward must eventually enter the Stokes regime
(also see Fig.~1 by \citet{LJ12}).

\subsection{Basic relation for the pebble accretion rate}

The mass accretion rate of pebbles onto a planetary embryo with mass
$M$ depends on whether the accretion may be considered as
bidimensional (if the scale height of the pebble disk is small
compared to the cross section of the collisions) or
three-dimensional. In the first case, the accretion rate is
\begin{equation}
\dot{M}_{\rm 2D} = 2b \sigp \Delta \v,
\label{eq:acc_rate2D}
\end{equation}
where $2b$ is the linear cross section of the collision ($b=R$ in the
geometric limit), $\hp$ and $\sigp$ are the scale height and 
the surface density of a pebble subdisk, and $\Delta \v$ is 
the relative velocity between the embryo and the pebbles. 

Otherwise, the situation becomes more complex. A limiting case is when
the gravitational pull of the embryo is large enough that the
pebble flux can be considered as isotropic. In that case, the
accretion rate can be written as 
\begin{equation}
\dot{M}_{\rm 3D}  =\pi b^2 \frac{\sigp}{\sqrt{2\pi} \hp}  \Delta \v,
\label{eq:acc_rate3D}
\end{equation}
where we used the fact that the pebble spatial density at the
midplane is given by $\rho_p \simeq \sigp/\sqrt{2\pi} \hp$.
The above equations can be combined into \citep[see also][]{Guillot14}
\begin{equation}
\dot{M} = \min\left(\sqrt{\frac{8}{\pi}} \frac{\hp}{b}, 1 \right) 
                  \times \sqrt{\frac{\pi}{2}} \frac{b^2}{\hp} \sigp \Delta \v.
\label{eq:acc_rate}
\end{equation}

The pebble scale height is related to the disk gas scale height as
\citep{Dubrulle95,Youdin_Lithwick07,Okuzumi12}
\begin{equation}
\hp \simeq \left(1+\frac{\st}{\alpha}\right)^{-1/2} \hg \simeq \left(\frac{\st}{\alpha}\right)^{-1/2} \hg,
\label{eq:h_p}
\end{equation}
where we assumed $\st/\alpha > 1$, because 
for pebble accretion, we usually consider the parameter ranges of
$\st \ga 0.1$ and $\alpha \la 10^{-2}$.
Because $\hg \propto r$, the accretion mode tends to be 2D
in the inner disk regions (also see section 4). 
In late phases, when planetary mass becomes so high that $b \ga \hp$, 
the accretion mode also changes from 3D to 2D (section 3.5). 
The transitional planetary mass is 
$O(10^{-1})\,\mearth$, as shown in Eq.~(\ref{eq:M2D3D}).

The radial and azimuthal components of pebble drift velocity are given by 
\citep[e.g.,][]{Nakagawa86,Guillot14}
\begin{equation}
\v_r = -\Lambda^2 \frac{2\st}{1+\Lambda^2 \st^2}\eta \vk
+ \frac{1}{1+\st^2}u_\nu,
\label{eq:vr}
\end{equation}
\begin{equation}
\v_\phi = -\Lambda \frac{1}{1+\Lambda^2 \st^2}\eta \vk
+ \frac{\st}{2(1+\st^2)} u_\nu,
\label{eq:vphi}
\end{equation}
where $\eta$ is the difference between gas and Keplerian velocities due to pressure gradient
given by $ (\hhg^2/2) |d \ln P/d \ln r|$,  
$\Lambda = \rho_g/(\rho_g + \rho_p)$,
and $u_\nu$ is the radial viscous diffusion velocity ($\sim - \nu/r \sim 
-\alpha \hg^2 \Omega/r \sim -\alpha \hhg^2 \vk$).
We hereafter assume $\Lambda \simeq 1$.
Because $\eta \sim \hhg^2$ and we consider the cases with $\alpha \la 10^{-2}$ and $\st \ga 0.1$,
we neglect the 2nd terms proportional to $u_\nu$ 
We note that when pebbles migrate to the region inside the snow line and are broken apart into small silicate grains, the second terms in Eqs.~\eqref{eq:vr} and \eqref{eq:vphi} could become important.
Since $\eta$ regulates pebble migration speed, it is a very important
quantity for pebble accretion.
It is explicitly given by
\begin{equation}
\eta = \frac{\hhg^2}{2} \left| \frac{d \ln P}{d \ln r} \right|
\simeq \left\{
\begin{array}{l}
0.93 \times 10^{-3} M_{*0}^{-7/10}
\alpha_3^{-1/5}
\dot{M}_{*8}^{2/5}
\left(\frac{r}{1\,\au}\right)^{2q} \\
\hspace*{2cm} [\mbox{viscous regime; } q = 1/20], \\
0.80 \times 10^{-3}
 L_{*0}^{2/7}M_{*0}^{-8/7} 
 \left(\frac{r}{1\,\au}\right)^{2q}  \\
 \hspace*{2cm} [\mbox{irradiation regime; } q = 2/7],
\end{array}
\right.
\label{eq:eta}
\end{equation}
where we used $d\ln P/d\ln r = d\ln (\sigg T/\hg)/d\ln r \simeq -2.55$
for the viscous region and $\simeq -2.78$ for the irradiation region.

Equations~(\ref{eq:vr}) and (\ref{eq:vphi}) are
rewritten as 
\begin{equation}
\v_r = - 2 \st \zeta \eta \vk,
\label{eq:zeta0}
\end{equation}
\begin{equation}
\Delta \v_0 \equiv \sqrt{\v_r^2 + \v_\phi^2} = \chi \eta \vk = \eta' \vk,
\label{eq:Delta_v}
\end{equation}
where 
\begin{equation}
\eta' = \chi \eta, \; \chi = \frac{\sqrt{1+ 4\st^2}}{1+\st^2}, \; \zeta = \frac{1}{1+\st^2}.
\label{eq:zeta}
\end{equation}
For $\st < 1$, $\zeta, \chi \simeq 1$, while $\zeta \simeq 1/\st^2$ and
$\chi \simeq 2/\st$ for $\st \gg 1$.
As we show in what follows, the dependence on $\chi$ disappears in 
pebble accretion rates, while the $\zeta$-factor remains.
It is no problem to assume $\chi = 1$ and $\eta' = \eta$ in the following.

If a circular orbit is assumed for the embryo, the relative velocity 
between the embryo and a pebble is given by the sum of their relative velocities $\Delta v_0$ and of a contribution due to the Keplerian shear \citep{Ormel_Klahr10,Guillot14}
\begin{equation}
\Delta \v \sim \eta' \vk + \frac{3}{2}b\Omega
  \sim \left(1 + \frac{3b}{2\eta' r}\right)\eta' \vk.
\label{eq:Dv}
\end{equation}

Departures from this assumption occur if the eccentricities or inclinations of the embryos become larger than $\eta \sim O(10^{-3})$ \citep{Guillot14}.
Density fluctuations from turbulence could excite eccentricities to  
$3 \times 10^{-3} \alpha_3^{1/3} (R/100{\rm km})^{1/3}(r/1\,\au)^{11/12}$ 
for $R\la 1000$ km \citep{Guillot14}. 
The collision velocity could be dominated by the embryo eccentricity
with $R \sim 100$--1000 km in the case of $\alpha \ga 10^{-3}$.
However, after $R$ becomes larger than 1000 km, 
disk-planet interaction efficiently damps eccentricity and 
the second term of Eq.~(\ref{eq:Dv}) becomes larger
than the first term (a transition from Bondi (drift accretion) regime to Hill regime; see below).
As a result, the effect of the embryo eccentricity becomes weak again.
The self-stirring of the embryos could also become important. Since it depends on orbital separation, which has not been clarified in the case of their formation by streaming instability, this is difficult to estimate at this point. 
We leave analysis of this case for future work and assume here for simplicity that
$\Delta \v$ is given by Eq.~(\ref{eq:Dv}).
In Eqs.~(\ref{eq:Delta_v}) and \eqref{eq:Dv}, the relative velocity induced by turbulence, $\sqrt{\alpha} c_s$, 
is not included. 
It is negligible for pebble size bodies, as long as $\alpha \la 10^{-2}$ \citep[e.g.,][]{Sato15}.

\citet{Moriarty15} used $\Delta \v=\v_r$, while $\v_r$ is smaller than
$\v_\theta$ for $\st < 1$ and the shear velocity is more important 
for high mass planets.
Furthermore, although they discussed formation of close-in planets,
they assumed irradiative scale height.
In general, viscous heating is more important than the stellar irradiation
in inner disk regions.  

The set of equations \eqref{eq:acc_rate}, \eqref{eq:h_p},
\eqref{eq:eta}, \eqref{eq:zeta} and \eqref{eq:Dv}
allows us to derive mass accretion rates for the different cases
that we consider.

\subsection{Cross section of pebble accretion}

We now need to calculate the accretion cross section $b$. 
Here, we consider 1-100 cm-sized pebbles accreted by a planetary embryo
with a size larger than 100 km. 
The gas drag effect is then combined with the gravitational pull of the embryo,
which results in a significant increase of the collision cross section
\citep[settling regime;][]{Ormel_Klahr10, Guillot14}.
For $\st < 1$, the velocity change of a pebble with an impact parameter $b$
by the gravitational force from the embryo
is given approximately by \citep{LJ12,Guillot14}
\begin{equation}
\delta \v \sim t_{\rm stop} \frac{GM}{b^2} 
\sim \st \frac{GM}{b^2 \Omega_K}.
\end{equation}
If $\delta \v \la \Delta \v/4$, a collision occurs \citep{Ormel_Klahr10}.
Then,
\begin{equation}
b^2 \sim \frac{4 \st GM}{\Omega \Delta \v}. 
\end{equation}

When $b \la (2/3)\eta' r$ and hence $\Delta \v \sim \eta' \vk$ (Bondi regime),
the equivalent radius cross section is 
\begin{equation}
b \simeq \left(\frac{12 \st \RH^3}{\eta' r}\right)^{1/2}
\simeq 2 \sqrt{\frac{t_{\rm stop}}{t_B}} R_B \; \; \; [{\rm Bondi \; regime}],
\label{eq:b_Bondi}
\end{equation}
where $R_B = GM/(\eta' \vk)^2$ and $t_B = R_B/\eta' \vk$
[$t_{\rm stop}/t_B \simeq \st \eta'^3 (M_*/M)$].
When $b \ga (2/3)\eta' r$ and hence $\Delta \v \sim (3/2) b\Omega$ (Hill regime),
\begin{equation}
b \simeq 2 \st^{1/3} \RH \simeq \left(\frac{\st}{0.1}\right)^{1/3} \RH
\; \; \; [{\rm Hill\; regime}].
\label{eq:b_Hill}
\end{equation}
This expression assumes $\st \la 0.1$. As $\st$
increases over this value, $b$ asymptotes to $\sim \RH$. 
However, while almost all trajectories with impact parameters $< b$
result in collisions with the planetary embryo for $\st \la 1$,
only a small fraction of
the trajectories with impact parameters within $\sim \RH$ can actually collide
for $\st \gg 1$  \citep[e.g.,][]{Ida_Nakazawa89, Ormel_Klahr10} 
because their motions are not dissipative.
 To take this effect into account, \citet{Ormel_Kobayashi12}
 proposed a reduction factor for $b$ with $\st \gg 1$ as
\begin{equation}
\kappa =\exp \left(- \left(\frac{\st}{\min(2,\st^*)}\right)^{0.65}\right),
\label{eq:kappa}
\end{equation}
where $\st^* = 4(M/M_*)/\eta^3$.

Taking the reduction factor into account,
Equations (\ref{eq:b_Bondi}) and (\ref{eq:b_Hill}) are combined into
\begin{equation}
b \simeq  \min\left(\sqrt{\frac{3 \st^{1/3} \RH}{\eta' r}}, 1 \right) \times 2 \kappa \st^{1/3} \RH.
\end{equation}
The left-hand term in the bracket dominates for small $M$ (Bondi regime).
The transition from the Bondi regime to Hill regime
(when the right-hand term in the bracket becomes comparable to
the left-hand term)
occurs at $b \sim (2/3)\eta' r$, which is equivalent to $\RH \sim (\eta'/3\st^{1/3}) r$.
The transition planetary mass is given by 
\begin{equation}
\begin{array}{ll}
M_{\rm BH} & 
{\displaystyle = \frac{\eta'^3}{9\st}  M_* \sim 10^{-9} \left(\frac{\st}{0.1}\right)^{-1} 
 \left(\frac{\eta}{10^{-3}}\right)^3 M_* } \\
 & {\displaystyle \sim 3 \times 10^{-4} \left(\frac{\st}{0.1}\right)^{-1} 
 \left(\frac{\eta}{10^{-3}}\right)^3  
 M_{\oplus}.}
\end{array}
\label{eq:M_BH}
\end{equation}
In the last two equations, we assumed $\st < 1$.

\subsection{Pebble mass flux and surface density}
\label{subsec:mig_pebble}

 To estimate the pebble mass flux, we first evaluate the size of
 migrating pebbles from the balance between growth and migration.
The  dust growth timescale is approximately given by
\citep{Takeuchi_Lin05,Brauer08}
 \begin{equation}
 t_{\rm grow} \simeq \frac{4}{\sqrt{3\pi}} \frac{\sigg}{\sigp} \Omega^{-1} 
 \simeq 20 \left(\frac{\sigp/\sigg}{10^{-2}}\right)^{-1} 
 \left(\frac{r}{1\,\au}\right)^{3/2}{\rm yrs},
 \label{eq:t_grow}
 \end{equation}
where we assumed perfect sticking for simplicity. 
For high enough speed collisions, grains rebound or fragment
rather than coagulate, which is called a bouncing or fragmentation barrier,
and  the threshold velocity may be 20-100 m/s for icy grains and 
about ten times smaller for silicate grains 
 \citep{Blum00, Zsom+2010,Zsom11,Wada11,Weidling12,Wada13}.
We show in the following that the barrier does not affect the assumption of perfect sticking for icy grains.
On the other hand, the barrier may prevent silicate grains from growing beyond millimeter sizes
\citep{Zsom+2010,Zsom11,Wada11,Weidling12}.
We will also discuss this issue later.

 Although Eq.~\eqref{eq:t_grow} differs slightly from more detailed calculations,
 for example, in the Stokes regime
 \citep{Sato15}, it is useful for the purpose of the present paper. 
In the region of pebble formation from sub-micron dust grains,
migration has not set in and we use a typical value $\sigp/\sigg
\sim 10^{-2}$. 
In regions where pebbles are migrating, 
$\sigp/\sigg$ becomes much smaller than $10^{-2}$ (see below).

The timescale of radial migration of dust due to gas drag is given by
 \begin{equation}
 \begin{array}{ll}
 t_{\rm mig} &
 {\displaystyle  = \frac{r}{\v_r} \simeq \frac{1+\st^2}{2\st}\frac{r}{\eta \vk}
 = \frac{1+\st^2}{\st} \frac{1}{2\eta\Omega} }\\
% {\displaystyle  = \frac{r}{\v_r} \simeq \frac{r}{2\st \eta \vk}
% = \frac{T_K}{4\pi \st \eta} }\\
 & {\displaystyle  
% \simeq 57 \frac{1+\st^2}{\st} T_{280}^{-1}\left(\frac{r}{1\,\au}\right)^{3/2-2q}
 \sim 0.8 \times 10^3 \left(\frac{\st}{0.1}\right)^{-1}
\left(\frac{\eta}{10^{-3}}\right)^{-1} {\rm yrs},
}
\end{array}
\label{eq:t_mig}
 \end{equation}
where we assumed $\st < 1$, because we consider relatively outer regions
for the formation site of pebbles.
 
For small dust grains with $\st \ll 1$, 
$t_{\rm mig}$ is much longer than $t_{\rm grow}$,
so that they grow without significant migration.
As pebbles grow, $t_{\rm mig}$ decreases, while $t_{\rm grow}$ does not change.
Since pebble growth occurs in an inside-out manner (Eq.~(\ref{eq:t_grow}))
and the largest bodies dominate the total pebble surface density \citep{Sato15},
$\sigp/\sigg$ can be regarded as a constant behind the pebble formation front.
Pebbles start their migration when $t_{\rm mig}$ becomes shorter than $t_{\rm grow}$.
That is, migration starts when $\st$ exceeds
 \begin{equation}
\tau_{\rm s, crit1} \sim
\frac{\sqrt{3\pi}}{8\eta}\frac{\sigp}{\sigg}.
\label{eq:tau_s_crit1}
 \end{equation}
 
The surface densities of migrating pebbles 
and disk gas are given by
\begin{equation}
\begin{array}{ll}
\sigp & 
{\displaystyle = \frac{\dot{M}_{\rm F}}{2\pi r \v_r} 
\sim \frac{1+\st^2}{\st}\frac{\dot{M}_{\rm F}}{4\pi r \eta \vk} },\\
 \sigg & 
 {\displaystyle \simeq \frac{\dot{M}_*}{3\pi \alpha \hg^2 \Omega},}
\label{eq:Sigma_p}
\end{array}
\end{equation}
where $\dot{M}_{\rm F}$ is the pebble mass flux through the disk.
Since $\eta = (1/2)(\hg/r)^2|d\ln P/d\ln r|$,
\begin{equation}
\begin{array}{ll}
{\displaystyle \frac{\sigp}{\sigg}} & 
{\displaystyle \simeq 
\frac{3}{2 | d\ln P/d\ln r |}\frac{1+\st^2}{\st}
\alpha
\frac{\dot{M}_{\rm F}}{\dot{M}_*}}\\
 & {\displaystyle \simeq 2 \times 10^{-4}
\left(\frac{\st/(1+\st^2)}{0.1}\right)^{-1} 
 \alpha_3 \dot{M}_{\rm F4}
\dot{M}_{*8}^{-1}.}
\end{array}
\label{eq:solid_to_gas}
\end{equation}
If we use an expression of $\dot{M}_{\rm F}$ given by Eq.~(\ref{eq:M_F_est}),
\begin{equation}
\frac{\sigp}{\sigg} \simeq 
2 \times 10^{-3} L_{*0}^{-2/7}M_{*0}^{9/14} 
\left(\frac{\st/(1+\st^2)}{0.1}\right)^{-1}
\left(\frac{t}{10^6{\rm yr}}\right)^{-1/3},
 \end{equation}
 which is consistent with the results by \citet{Sato15} and \citet{Krijt16}.

The parameter $\sigp$ is smaller than the solid surface density of the MMSN
by two orders of magnitude for $\st \sim 0.1$ and
$\dot{M}_{\rm F} \sim 10^{-4}\,\mearth/{\rm yr}$, which was
also found by \citet{LJ14b}.
This is because the pebble migration velocity $\v_r$ is very large 
and we consider a steady-state solution in which $\dot{M}_{\rm F}$ is
independent of $r$.
If pebbles split into mm-sized silicate grains when they pass through the snow line
as assumed by \citet{Morbidelli15a}, we expect $\sigp$ to increase
again inside the snow line and to become comparable to the MMSN value. 

Substituting Eq.~(\ref{eq:solid_to_gas})
into Eq.~(\ref{eq:tau_s_crit1}) with $\st \sim \tau_{\rm s,crit1}$
and assuming $\st < 1$, we obtain 
\begin{equation}
\begin{array}{ll}
\tau_{\rm s, crit1} &
{\displaystyle
\simeq \left( \frac{3\sqrt{3 \pi}}{16 |d\ln P/d\ln r|}\frac{\alpha}{\eta}\frac{\dot{M}_{\rm F}}{\dot{M}_*} \right)^{1/2}
} \\
 & {\displaystyle \simeq 0.08 
\left(\alpha_3 \dot{M}_{*8}^{-1} \dot{M}_{\rm F4} \right)^{1/2}
\left(\frac{\eta}{10^{-3}}\right)^{-1/2}  \;  \propto r^{-q}.} \\
& {\displaystyle \rightarrow 0.25  L_{*0}^{1/7}M_{*0}^{-9/28} \left(\frac{\eta}{10^{-3}}\right)^{-1/2}}
 \; \; [\mbox{if Eq.~(\ref{eq:M_F_est}) is assumed}],
\end{array}
\label{eq:tau_s_crit1B}
 \end{equation}

In the Epstein regime, $\st \propto R \, r^{\xi}$ (Eq.~(\ref{eq:tau_s})), 
where $\sigg \propto r^{-\xi}$.
Both $q$ and $\xi$ are usually positive.
As a pebble migrates inward, without growth, its $\st$ would
decrease. However, because $\tau_{\rm s, crit1}$ increases,
growth must dominate over migration. 
Here we consider collisions between pebbles and
used Eq.~(\ref{eq:solid_to_gas}) for $\tau_{\rm s,crit1}$.
This means that pebbles must migrate and grow 
so that $\st \sim \tau_{\rm s,crit1}$
(see Fig.~\ref{fig:R_evol} below).
In the irradiation regime, the pebble size evolves with $\st \sim \tau_{\rm s,crit1}$ 
(Eq.~(\ref{eq:tau_s3})) as
\begin{equation}
\begin{array}{l}
{\displaystyle 
R  \simeq 87 L_{*0}^{-3/7} M_{*0}^{17/14}
  \alpha_3^{-1/2}
  \dot{M}_{*8}^{1/2}
  \dot{M}_{\rm F4}^{1/2}
  \rho_{s1}^{-1} \left(\frac{r}{1\,\au}\right)^{-19/14}{\rm cm},}
\end{array}
\label{eq:pebble_size}
\end{equation}
where we used Eq.~(\ref{eq:eta}) in the irradiation region.
 
As pebbles further grow and migrate inward, $\lambda_{\rm mfp}$
becomes smaller until they eventually enter the Stokes regime when $R \ga (9/4)\lambda_{\rm mfp}$.
Equations~(\ref{eq:l_g}) and (\ref{eq:pebble_size}) show that
the Epstein-Stokes transition in the irradiation region occurs at 
\begin{equation}
r  \simeq r_{\rm ES} = 
 2.9 L_{*0}^{-3/13} M_{*0}^{17/26}  
  \alpha_3^{-21/52}
  \dot{M}_{*8}^{21/52}
  \dot{M}_{\rm F4}^{7/52}
  \rho_{s1}^{-7/26} \au.
\label{eq:Ep_St}
\end{equation}
Because  $\st \propto R^2 \times r^{-1-q}$ in the Stokes regime,
$\st$ increases with inward migration and 
$\st > \tau_{\rm s, crit1}$ is always satisfied.
In the Stokes regime, pebbles migrate without significant growth.
Therefore, the pebble size in the Stokes regime is given by
 substituting Eq.~(\ref{eq:Ep_St}) into Eq.~(\ref{eq:pebble_size}) as
\begin{equation}
 R \simeq 
 20 L_{*0}^{-0.11} M_{*0}^{0.33} 
 \alpha_{3}^{0.05}  \dot{M}_{*8}^{-0.05}
  \dot{M}_{\rm F4}^{0.32}
  \rho_{s1}^{-0.63} {\rm cm}.
\label{eq:Ep_St_R}
\end{equation}
The corresponding evolution of the Stokes number is obtained
by substituting Eq.~(\ref{eq:Ep_St_R}) into Eq.~(\ref{eq:tau_s2}).

With a constant $R$, the Stokes parameter increases with inward migration
as $d\ln \st/d \ln r \simeq -1-q$ in the Stokes regime. 
When $\st$ exceeds unity, 
the migration quickly slows down 
($t_{\rm mig} \propto \st$ for  $\st > 1$).
If $\st$ would further exceed 
\begin{equation}
\begin{array}{ll}
 \tau_{\rm s, crit2} &  
 {\displaystyle
\simeq \left( \frac{3\sqrt{3 \pi}}{16 |d\ln P/d\ln r|}\frac{\alpha}{\eta}\frac{\dot{M}_{\rm F}}{\dot{M}_*} \right)^{-1/2}
} \\
 & {\displaystyle \simeq 13  \left(\alpha_3^{-1} \dot{M}_{*8} \dot{M}_{\rm F4}^{-1} \right)^{1/2}
 \left(\frac{\eta}{10^{-3}}\right)^{1/2}} \;\; \propto r^q\\
 & {\displaystyle \rightarrow 4  L_{*0}^{1/7}M_{*0}^{-9/28} \left(\frac{\eta}{10^{-3}}\right)^{1/2}}
 \; \; [\mbox{if Eq.~(\ref{eq:M_F_est}) is assumed}],
 \end{array}
 \label{eq:tau_s_crit2}
 \end{equation}
before the pebbles pass the snow line,
$t_{\rm mig}$ would again become longer than $t_{\rm grow}$
and the pebbles would keep growing {\it in situ}.
Furthermore, since $\st \propto R^2$ , the condition $\st >
\tau_{\rm s, crit2}$ would then always be satisfied, resulting in runaway coagulation \citep{Okuzumi12}.
\citet{Okuzumi12}, however, showed that $\st$ does not
reach $\tau_{\rm s, crit2}$ outside the snow line where
the assumption of perfect accretion may be relevant,
unless we consider the possibility that dust grains are highly
porous. We do not consider this possibility in the present work. 

So far, we assumed perfect accretion.
However, even for icy grains, the bouncing/fragmentation barrier exists as mentioned before.
For collisions between dust grains, their collision velocity is likely to be
dominated by that induced by turbulence as long as $\alpha \ga 10^{-3}$,
which is given by 
$\v_{\rm col} \sim \sqrt{3 \alpha \st} c_s$ \citep{Sato15}.
For the threshold velocity $\v_{\rm col,crit}$,
collisions result in coagulation when
\begin{equation}
\st < \frac{1}{3\alpha}\left(\frac{\v_{\rm col,crit}}{c_s}\right)^2
\simeq 2 \alpha_3^{-1} \left(\frac{\v_{\rm col,crit}}{50{\rm \,m/s}}\right)^2
\left(\frac{T}{150{\rm \,K}}\right)^{-1}.
\end{equation}
If we use $T_{\rm irr}$ given by Eq.~(\ref{eq:T_irr}),
\begin{equation}
\st \la 2 \alpha_3^{-1} \left(\frac{\v_{\rm col,crit}}{50{\rm \,m/s}}\right)^2
L_{*0}^{-2/7}M_{*0}^{1/7} \left(\frac{r}{1\,\au}\right)^{3/7}.
\end{equation}
Since collisions are efficient in Epstein regime, the result in Figure~1 shows that 
the bouncing/fragmentation barrier does not restrict the pebble growth
as long as $\alpha \la 10^{-2}$. 

\begin{figure}[htb]
\includegraphics[width=75mm,angle=0]{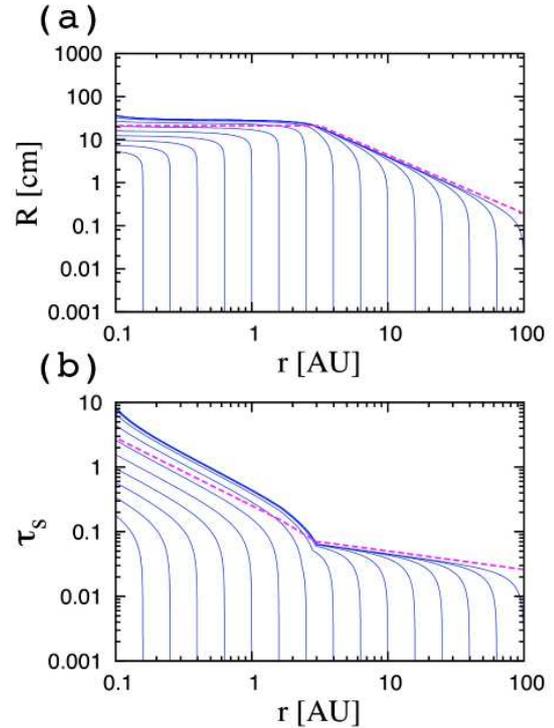}
\caption{Evolution of (a) size and (b) Stokes number
of pebbles migrating from gas drag for
$\dot{M}_{*8}=\dot{M}_{\rm F4}=\alpha_3=1$.
The dust grains are initially 0.001 cm in size and their 
growth and migration paths are calculated by directly integrating $dR/dt = R/t_{\rm grow}$
and $dr/dt = - r/t_{\rm mig}$ with Eqs.~(\ref{eq:t_grow}) and (\ref{eq:t_mig}) as bold lines.
The dashed lines represent the analytical estimates of
Eqs.~(\ref{eq:pebble_size}) and (\ref{eq:Ep_St_R}) in panel a,
and those in panel b are $\tau_{\rm s,crit1}$ (Eq.~(\ref{eq:tau_s_crit1B})) and
 Eq.~(\ref{eq:tau_s2}) with Eq.~(\ref{eq:Ep_St_R}).
}
\label{fig:R_evol}
\end{figure}
Figure~\ref{fig:R_evol}a shows the growth and migration of pebbles
obtained by directly integrating $dR/dt = R/t_{\rm grow}$
and $dr/dt = - r/t_{\rm mig}$ with Eqs.~(\ref{eq:t_grow}) and (\ref{eq:t_mig}).
In panel b, the corresponding evolution of $\st$ is plotted.
The analytical estimates are represented by the dashed lines, which are
given by
Eq.~(\ref{eq:pebble_size}) and $\tau_{\rm s,crit1}$ (Eq.~(\ref{eq:tau_s_crit1B})) 
in the Epstein regime and
Eq.~(\ref{eq:Ep_St_R}) and Eq.~(\ref{eq:tau_s2}) with Eq.~(\ref{eq:Ep_St_R})
in the Stokes regime.
They are consistent with the results by direct integration.
These results are also consistent with the evolution of
peak mass bodies obtained by more detailed dust growth/migration
calculations taking the dust size distribution  into account
\citep[e.g.,][]{Okuzumi12,Sato15}.
\def\hhg{\hat{h}_{\rm g}}
\citet{LJ14b} derived a similar analytical result.
However, they were focused on Epstein regime, so that
the results differ in inner disk regions where Stokes regime is important. 

From this plot, we find that runaway coagulation appears inside 0.1\au.
Indeed, the threshold $\tau_{\rm s,crit2}$ crosses the point of $r=0.1\au$ and
$\st=10$ with a positive gradient $q \simeq 1/20$.
However, the sublimation of icy components inside the snow
line would prevent runaway coagulation from occurring.

The pebble mass flux is evaluated as follows.
The pebble growth timescale ($t_{\rm grow}$) 
is the timescale for a body to grow in size by a factor of $e \sim 2.72$.
The timescale of growth from $\mu$m dust to cm pebbles (by a factor $10^4$)
is $t_{\rm p,grow} \sim \ln 10^4 \times t_{\rm grow} \sim 10 t_{\rm grow}$.
From Eq.~(\ref{eq:t_grow}), 
$t_{\rm p,grow} \sim  2 \times 10^5 (r/100\,\au)^{3/2}{\rm yrs}$,
where $\sigp$ is the small dust surface density as $\sigp/\sigg \sim 10^{-2}$.

We assume $\sigp/\sigg \sim 10^{-2}$ in the outer pebble-forming
region and use the value of $\sigg$ that applies to the irradiation regime (Eq.~(\ref{eq:Sigma_irr})).
The parameter $t_{\rm p,grow}$ depends on $\sigp/\sigg$ but not on $\sigp$.
Once migration starts, pebbles quickly migrate and the dust surface density there is rapidly depleted.
The timescale for pebbles to grow until migration dominates, $t_{\rm p,grow}$,
is proportional to $r^{3/2}$.
Thereby, the region in which pebbles are forming region is narrow
and migrates outward.
The pebble formation front $r_{\rm pf}$ at $t$ 
satisfies $t \sim 2 \times 10^5 (r_{\rm pf}/100\,\au)^{3/2}{\rm yrs}$,
that is, $r_{\rm pf} \sim 100 (t/2\times 10^5{\rm yr})^{2/3}\au$.
The pebble mass flux is estimated by calculating the dust mass
swept by the pebble formation front per unit time \citep{LJ14b},
\begin{equation}
\begin{array}{ll}
\dot{M}_{\rm F, \infty} & {\displaystyle \sim 2\pi r_{\rm pf} 
\times 0.01 \sigg(r_{\rm pf}) \times \frac{dr_{\rm pf}}{dt}}
\\
 & \simeq 9.4 \times 10^{-4}
L_{*0}^{-2/7}M_{*0}^{9/14} \alpha_3^{-1} \dot{M}_{*8}
 \left(\frac{t}{10^6{\rm yr}}\right)^{-1/3} \;\;\mearth/{\rm yr}.
 \end{array}
 \label{eq:M_F_est}
\end{equation}
The pebble mass flux $\dot{M}_{\rm F}$ is
governed by the outward migration of pebble formation front, 
but not by the pebble migration speed.

A disk with a surface density $\Sigma_{\rm g,irr}$ defined by Eq.~(\ref{eq:Sigma_irr}) can be gravitationally unstable in its outer regions. The Toomre Q parameter is given by
\begin{equation} Q = \frac{c_s\Omega}{\pi G\Sigma_{\rm g,irr}}\simeq  \hat{h}_{\rm g,irr}\frac{M_*}{\pi \Sigma_{\rm g,irr} r^2}   \simeq 25 L_{*0}^{2/7}M_{*0}^{-9/14} \alpha_3 \dot{M}_{*8}^{-1}  \left(\frac{r}{1\,\au}\right)^{-9/14}. \end{equation}
If the turbulence due to the disk instability is so vigorous that even icy grains do not grow, the pebble formation front is given by
\begin{equation} 
r_{\rm pf} \simeq 50  \left(\frac{Q}{2}\right)^{-14/9}  L_{*0}^{4/9}M_{*0} \alpha_3^{14/9} \dot{M}_{*8}^{-14/9}\, \au. 
\end{equation}
According to our $\alpha$ disk evolution model, $\dot{M}_{*8}$ decreases monotonously while $r_{\rm pf}$ increases. If $\dot{M}_{*8}\sim (t/10^6{\rm yr})^{-3/2}$, $r_{\rm pf} \sim 50 (t/10^6{\rm yr})^{7/3} \au.$ In this case, $\dot{M}_{\rm F}$ is 2 times smaller than that given by Eq.~(\ref{eq:M_F_est}).

In the above derivation, it is assumed that the disk is extended to infinity.
\citet{Sato15} showed the existence of
 two phases for the flux of pebbles (their Fig. 6).
 First, the pebble mass flux $\dot{M}_{\rm F}$ is almost constant with time until
$r_{\rm pf}$ exceeds the disk size $r_{\rm out}$.
After that, $\dot{M}_{\rm F}$ decays rapidly as a consequence of the depletion of solid materials in the outer regions.
The transition time between the two regimes is
\begin{equation}
t_{\rm peb. dep.} \sim 2 \times 10^5 (r_{\rm out}/100\,\au)^{3/2}{\rm yrs}.
\label{eq:tpebdep}
\end{equation}
If we assume an exponential taper to the surface density of the disk beyond $r_{\rm out}$, $\sigg \propto r^{-1} \exp(-r/r_{\rm out})$,
the two phase evolution is 
described by addition of a decaying factor of
$\exp(-r_{\rm pf}/r_{\rm out}) = \exp\left(-(100\,\au/r_{\rm out}) (t/2 \times 10^5  {\rm yr})^{2/3}\right)$
to $\dot{M}_{\rm F}$ given by of Eq.~(\ref{eq:M_F_est}).
Because observations suggest that disks around T Tauri stars typically have 
$r_{\rm out}\la 100 \,\au$ \citep[e.g.,][]{Andrews09},
the reduction cannot be neglected.

Furthermore, if the planetary embryos grow beyond Mars mass,
a non-negligible fraction of $\dot{M}_{\rm F}$ may be filtered out
by the accretion onto the embryos (see section~\ref{sec:acc_rate} hereafter).
Thus, $\dot{M}_{\rm F}$ and $\sigp$ would be quickly depleted
once such large embryos appear (see also Chambers 2016, submitted).

The time evolution of $\dot{M}_{\rm F}$ is very important
for the final configuration of planetary systems formed through pebble accretion.
However, a proper calculation of this quantity would be beyond the scope of the present work 
and we choose to treat $\dot{M}_{\rm F}$ as a constant parameter, adopting as a nominal value 
$\dot{M}_{\rm F}=10^{-4} \dot{M}_{*8}\; \mearth/{\rm yr}$. 
This value, obtained from dust 
growth calculations \citep{LJ12,Sato15}, is about an order of 
magnitude smaller than that obtained in Eq.~\eqref{eq:M_F_est}. We note that there is 
considerable uncertainty on this value which also depends on the disk surface density profile.

\subsection{General form of the pebble accretion rate}\label{sec:acc_rate}

We now seek to calculate the rate of pebble accretion by a protoplanet
and whether its presence can affect the flow of pebbles. From Eqs.~(\ref{eq:Dv}) and (\ref{eq:Sigma_p}),
\begin{equation}
\sigp \Delta \v \sim 
\frac{\dot{M}_{\rm F}}{4\pi r\st} \zeta^{-1} \chi \left(1 + \frac{3b}{2 \chi \eta r}\right).
\end{equation}
Substituting this into Eq.~(\ref{eq:acc_rate}), the pebble accretion rate is given by
\begin{equation}
\dot{M} = \frac{C \zeta^{-1} \chi \hat{b}^2}{4 \sqrt{2\pi} \st \hhp}\left(1 + \frac{3\hat{b}}{2\chi \eta}\right) \dot{M}_{\rm F}, 
\label{eq:pebble_acc_rate}
\end{equation}
with 
\begin{equation}
C = \min\left(\sqrt{\frac{8}{\pi}}\frac{\hp}{b},1 \right) =  \min\left(\sqrt{\frac{8}{\pi}} \frac{\hhp}{\hat{b}}, 1 \right),
\end{equation}
\begin{equation}
\hat{b} =\min\left(1, \sqrt{3 \st^{1/3} \hat{R}_H/\eta'}\right) \times 2 \kappa \st^{1/3} \hat{R}_H
\label{eq:b_hat}
\end{equation}
where the hat sign such as $\hat{b}$ expresses lengths 
that have been divided by the orbital distance $r$.
Here $\dot{M}_{\rm F}$ and $\st$ are given as parameters, such as 
$\dot{M}_{\rm F} \sim 10^{-4}M_{\oplus}$ and $\st \sim 0.1$
(see section \ref{subsec:mig_pebble}).
The formula for $\dot{M}$ given by Eq.~(\ref{eq:pebble_acc_rate})
can exceed $\dot{M}_{\rm F}$, in which case we limit it to that
value. 

From Eq.~(\ref{eq:pebble_acc_rate}), the accretion timescale is 
\begin{equation}
t_{\rm acc} = \frac{M}{\dot{M}} = \frac{M}{\dot{M}_{\rm F}} \frac{4 \sqrt{2\pi} \zeta \st \hhp}{C \chi \hat{b}^2}
 \left(1 + \frac{3\hat{b}}{2\chi\eta}\right)^{-1}.
 \label{eq:pebble_acc_time}
\end{equation}
In the 3D case ($C \sim 1$), the pebble accretion timescale ($t_{\rm acc,3D}$)
has an identical simple form both in the Bondi and Hill regimes, as shown below.
In the early Bondi phase ($\hat{R}_H \la \chi \eta/3\st^{1/3}$),
$\hat{b} \sim 2\sqrt{3} \kappa  (\chi \eta)^{1/2} \st^{1/2}  \hat{R}_H^{3/2}
\la 2\eta/3$, and Eq.~(\ref{eq:pebble_acc_time}) reads as 
\begin{equation}
\begin{array}{ll}
t_{\rm acc,3D} 
&
{\displaystyle 
 \simeq \frac{M}{\dot{M}_{\rm F}} \frac{4 \sqrt{2\pi} \zeta \st \hhp}{\chi \hat{b}^2} }\\
 & {\displaystyle
 \simeq \sqrt{2\pi} \frac{\zeta}{\kappa^2} \eta \hhp \frac{M_*}{\dot{M}_{\rm F}}
 \simeq \sqrt{2\pi} \frac{\zeta}{\kappa^2}  \eta \hhg \left(\frac{\alpha}{\st}\right)^{1/2}\frac{M_*}{\dot{M}_{\rm F}},}
 \end{array}
 \label{eq:t_acc_3D}
\end{equation}
where we used $\hhp \sim [1+(\st/\alpha)]^{-1/2}\hhg \sim (\alpha/\st)^{1/2} \hhg$.
In the late Hill phase in which $\hat{b} \sim 2 \kappa \st^{1/3} \hat{R}_H$
and $\hat{b} \ga 2\chi \eta/3$,
\begin{equation}
t_{\rm acc,3D} \simeq \frac{M}{\dot{M}_{\rm F}} \frac{4 \sqrt{2\pi} \zeta \st \hhp}{\chi\hat{b}^2} \frac{2\chi \eta}{3\hat{b}}
\simeq \sqrt{2\pi}\frac{\zeta}{\kappa^{3}} \eta \hhg \left(\frac{\alpha}{\st}\right)^{1/2}\frac{M_*}{\dot{M}_{\rm F}},
 \label{eq:t_acc_3D2}
\end{equation}
which is identical to Eq.~(\ref{eq:t_acc_3D})
except for the reduction factor for $\st \gg 1$, although $\hat{b}$ has a different form.
Because the reduction does not actually occur in the Bondi regime,
we can use Eq.~(\ref{eq:t_acc_3D2}) both in the Bondi and Hill regimes.
For $\st > 1$, both $\zeta$ and $\kappa^{3}$ 
decrease (Eqs.~(\ref{eq:zeta}) and (\ref{eq:kappa})). 
Since $\kappa$ is a stronger function of $\st$, 
$\zeta/\kappa^3$ rapidly increases with $\st$ so that pebble
accretion slows down for $\st \gg 1$.
 Using $\hhg \simeq (2\eta/|\frac{d \ln P}{d \ln r}|)^{1/2} \sim 0.9 \eta^{1/2}$
(Eq.~(\ref{eq:eta})) and assuming $\st < 1$,
\begin{equation}
t_{\rm acc,3D} \simeq 
2 \times 10^4 \alpha_3^{1/2} 
L_{0*}^{3/7} M_{0*}^{-5/7}
\dot{M}_{\rm F4}^{-1}
\left(\frac{\eta}{10^{-3}}\right)^{3/2} 
\left(\frac{\st}{0.1}\right)^{-1/2} \; {\rm yrs}
\; \; \propto r^{3q}. 
\label{eq:t3D}
\end{equation}

The accretion mode is initially 3D.
It becomes 2D when $\hat{b} > \sqrt{8/\pi} \;\hhp$.
In this case, since it is likely that a transition from the Bondi
regime to the Hill regime 
($\hat{b} \sim 2 \kappa \st^{1/3} \hat{R}_H$) has already occurred,
the transition to the 2D accretion mode occurs when $2 \kappa \st^{1/3} \hat{R}_H \sim \sqrt{8/\pi} \;\hhp$,
that is, when
\begin{equation}
\begin{array}{ll}
M & {\displaystyle 
\sim M_{\rm 2D3D} \equiv \frac{3 (\sqrt{2/\pi} \; \kappa^{-1} \hhp)^3}{\st} M_*
\simeq 1.5 \kappa^{-3} \left(\frac{\alpha}{\st}\right)^{3/2}\frac{\hhg^3}{\st}M_*} \\
   & {\displaystyle 
   \simeq 0.1 \alpha_3^{3/2} \kappa^{-3}
\left(\frac{\st}{0.1}\right)^{-5/2} 
\left(\frac{\eta}{10^{-3}}\right)^{3/2} \,\mearth \; \; \propto r^{3q}.}
\end{array}
\label{eq:M2D3D}
\end{equation}
The transition mass from the Bondi to the Hill regime,
$M_{\rm BH}$ given by Eq.~(\ref{eq:M_BH}),
is actually much smaller than $M_{\rm 2D3D}$.
2D accretion always takes place in the Hill regime.

The accretion timescale in 2D is given by
\begin{equation}
t_{\rm acc,2D} =  \sqrt{\frac{\pi}{8}} \frac{\hat{b}}{\hhp} t_{\rm acc,3D} 
\simeq \frac{\pi}{2} \frac{\zeta}{\kappa^2} \eta \hat{b} \frac{M_*}{\dot{M}_{\rm F}}.
\end{equation}
Since the 2D mode occurs in the Hill regime, 
\begin{equation}
\begin{array}{ll}
t_{\rm acc,2D} & 
{\displaystyle \simeq \pi  \st^{1/3} \frac{\zeta}{\kappa^2} \eta \left(\frac{M}{3M_*}\right)^{1/3} \frac{M_*}{\dot{M}_{\rm F}} }\\
  & 
{\displaystyle
\simeq 2 \times 10^4 
L_{0*}^{2/7}M_{0*}^{-10/21}
\dot{M}_{\rm F4}^{-1}
\left(\frac{\st}{0.1}\right)^{1/3}
\left(\frac{\eta}{10^{-3}}\right)  
\left(\frac{M}{0.1\,\mearth}\right)^{1/3}
{\rm yrs} \; \; \propto r^{2q}.
}
\end{array}
\label{eq:t_acc_2D}
\end{equation}
In the last equation, we assumed $\st < 1$.
This accretion timescale is consistent with 
that derived by \citet{LJ14b} (their Eq.~(31)).

Following \citet{Guillot14}, we define the filtering efficiency by the ratio of
the accretion rate onto the embryo ($\dot{M}$) to that of the supplied pebble mass flux ($\dot{M}_{\rm F}$).
For $\st < 1$, the filtering efficiency in the 3D case ($M \la 0.1\,\mearth$) is
\begin{equation}
\begin{array}{ll}
P_{\rm 3D} & 
{\displaystyle = \frac{\dot{M}}{\dot{M}_{\rm F}}=  \frac{M}{\dot{M}_{\rm F} \; t_{\rm acc,3D}} 
} \\
  & 
{\displaystyle \simeq 0.05 
\alpha_3^{-1/2}
L_{0*}^{-3/7}M_{0*}^{5/7}
\left(\frac{\eta}{10^{-3}}\right)^{-3/2} 
\left(\frac{\st}{0.1}\right)^{1/2} 
\left(\frac{M}{0.1\,\mearth}\right) \; \; \propto r^{-3q}.}
\end{array}
\label{eq:P3D}
\end{equation}
The reduction of the pebble mass flux due to the accretion by a protoplanet
is thus negligible until $M$ becomes comparable to the mass of Mars.
Because $P_{3D} \propto M$, the total reduction in pebble mass flux 
from the filtering depends only on the total mass of planetary embryos,
as long as we consider the viscous ($q \simeq 1/20$), 3D-settling regime.  
The filtering probability in the 2D case ($M \ga 0.1\,\mearth$) is
\begin{equation}
\begin{array}{ll}
P_{\rm 2D} & {\displaystyle = \frac{M}{\dot{M}_{\rm F} \; t_{\rm acc,2D}} }
 \\
 & {\displaystyle \simeq 0.05
 L_{0*}^{-2/7}M_{0*}^{10/21}
 \left(\frac{\eta}{10^{-3}}\right)^{-1} 
\left(\frac{\st}{0.1}\right)^{-1/3} 
 \left(\frac{M}{0.1\,\mearth}\right)^{2/3} \; \; \propto r^{-2q}.}
\end{array}
\label{eq:P2D2}
\end{equation}
These values of $P_{\rm 3D}$ and $P_{\rm 2D}$ coincide with
those obtained by more detailed calculations by \citet{Guillot14}.

Substituting $M_{\rm 3D2D}$ given by  Eq.~(\ref{eq:M2D3D}) into Eq.~(\ref{eq:P3D}) or Eq.~(\ref{eq:P2D2}),
we find that the 2D probability must be applied when 
\begin{equation}
P > P_{\rm max} \simeq 0.05
\alpha_3  \left(\frac{\st}{0.1}\right)^{-2},
\end{equation}
which is independent of $\eta$.
For $P > P_{\rm max}$, $P$ increases proportionally to $M^{2/3}$
rather than to $M$: The total filtering by all the embryos becomes less efficient as they grow.

\section{Radial dependence of the pebble accretion timescale}

\begin{figure*}[htb]
\includegraphics[width=170mm,angle=0]{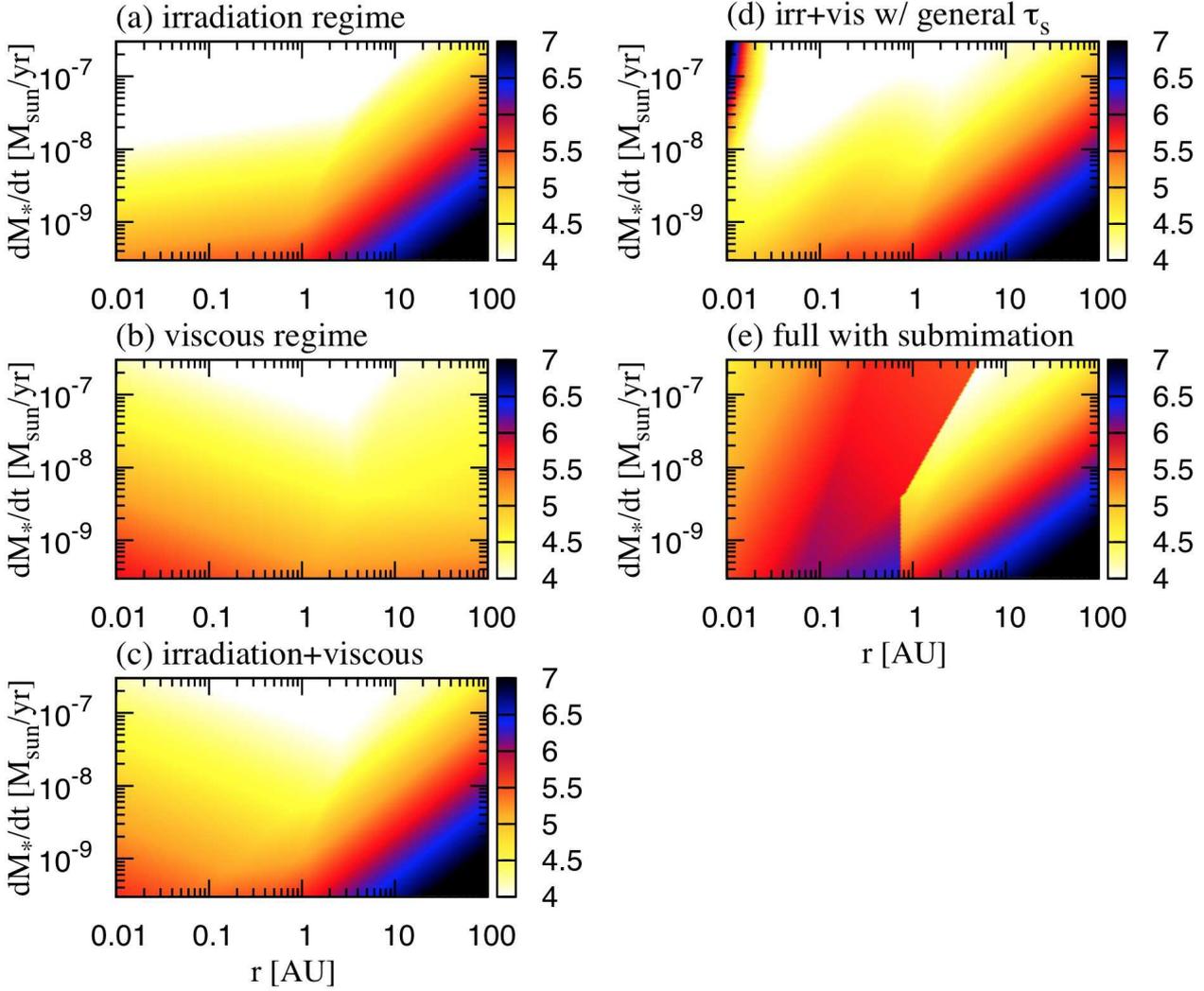}
\caption{Pebble accretion timescale $t_{\rm acc}$ of embryos
with $M=0.1\,\mearth$ on the $r$-$\dot{M}_*$ plane.
Color bars represent $\log_{10} (t_{\rm acc}/{\rm yr})$.
Brighter parameter regions represent faster accretion.
In panel a, the entire disk is assumed to be irradiation-heated, 
while it is viscously-heated in panel b. 
In panel c, the disk is the combination of irradiation-heated and viscously heated regimes.
In panel d, the reduction factor for $\st>1$ is taken into account in the disk in panel c.
In panel e, the effect of ice sublimation is added to the disk in panel d.
Details of individual sets are described in the main text.  
}
\label{fig:acc01} 
\end{figure*}

We consider the $r$-dependence of the pebble accretion timescale
that controls the final configuration of planetary systems.
We use $\st < 1$, $\hhg \propto r^q$ ($\eta \propto r^{2q}$), $\st \propto r^p$,
and $\sigg \propto r^{-\xi}$.
In the viscous region, $q \simeq 1/20$ and $\xi = 3/5$
and in the irradiation region $q \simeq 2/7$ and $\xi = 15/14$.
The exponent $p$ ($\st \propto r^p$) must be treated carefully.
In the Epstein regime, pebbles grow by keeping $\st \simeq \tau_{\rm s,crit1}$.
From Eq.~(\ref{eq:tau_s_crit1B}), $p \simeq -q$.
In the Stokes regime, pebbles migrate without significant growth, 
so that from Eq.~(\ref{eq:tau_s}), $p \simeq -1- q$.

From Eqs.~(\ref{eq:t_acc_3D}) and (\ref{eq:t_acc_2D}),
\begin{equation}
t_{\rm acc,3D} \propto  \frac{M^0}{\st^{1/2}} r^{3q} \propto 
M^0  r^{3q-p/2},
\label{eq:tacc3D_rdep}
\end{equation}
\begin{equation}
t_{\rm acc,2D} \propto M^{1/3} \st^{1/3} r^{2q} \propto 
M^{1/3} r^{2q+p/3},
\label{eq:tau_rdep_2D}
\end{equation}
where we also explicitly included the dependences on the planet mass ($M$) and 
Stokes parameter ($\st$).

The 3D accretion timescale is independent of $M$, which means
that planet growth is exponential.
In the case of planetesimal accretion,
the early runaway growth is superexponential: $t_{\rm acc} \propto M^{-1/3}$,
while the late oligarchic growth is subexponential: $t_{\rm acc} \propto M^{1/3}$
\citep[e.g.,][]{Kokubo_Ida98,Kokubo_Ida02}.
If conventional km-sized planetesimals are successfully formed, planetesimal accretion would dominate the early phases.
Pebble accretion would dominate in the oligarchic growth stage.
Pebble accretion then eventually enters the 2D mode and becomes comparable to
planetesimal accretion. However the two forms differ in the sense that
embryos can become isolated from planetesimals after depleting their
feeding zones \citep[e.g.,][]{Kokubo_Ida98,Kokubo_Ida02}, 
a process that would not occur in the case of pebbles
at least until the planet mass becomes high enough to
create a density gap in the disk that may halt the inward migration of pebbles 
and their supply to the embryos \citep[e.g.,][]{LJ12}.
This mass (pebble isolation mass) is comparable to
the inferred core masses of Jupiter and Saturn.

In the case of $\st < 1$, which is valid except in close-in regions,
the exponent ($t_{\rm acc} \propto r^\delta$) in the early 3D phase is:
\begin{equation}
\delta = 3q-p/2 \simeq
\left\{
\begin{array}{ll}
7q/2  = 7/40  & [\mbox{viscous \& Epstein}], \\
7q/2  = 1  & [\mbox{irradiative \& Epstein}], \\
7q/2 + 1/2  = 27/40 & [\mbox{viscous \& Stokes}], \\
7q/2 + 1/2  = 3/2 & [\mbox{irradiative \& Stokes}] .
\end{array}
\right.
\end{equation}
When embryos sufficiently grow or are located in sufficiently inner disk regions,
$M > M_{\rm 2D3D}$  (Eq.~(\ref{eq:M2D3D})) is satisfied and 
pebble accretion enters a 2D mode.
The exponent is
\begin{equation}
\delta = 2q+p/3 \simeq
\left\{
\begin{array}{ll}
5q/3 = 1/12 & [\mbox{viscous \& Epstein}], \\
5q/3= 15/21  & [\mbox{irradiative \& Epstein}], \\
5q/3 - 1/3 = -1/4  & [\mbox{viscous \& Stokes}],\\
5q/3 - 1/3 = 1/21  & [\mbox{irradiative \& Stokes}].
\end{array}
\right.
\end{equation}
When $\delta < 0$, the outer planets grow more rapidly.
Conversely, when $\delta > 0$, the inner planets grow more rapidly
(unless there is a significant reduction of the pebble flux from
filtering by outer planets). 

Pebble accretion hence has a weak dependence on $r$, but the fact that
$\delta$ can be either positive or negative has strong implications
for our understanding of planet formation. This is a very different
situation than the classical 
planetesimal accretion scenario for which the timescale strongly depends on $r$ 
($\propto 1/\sigp \Omega \propto r^{\xi + 3/2}$), and planetary
growth thus proceeds in an inside-out manner.

\begin{figure}[htb]
\includegraphics[width=80mm,angle=0]{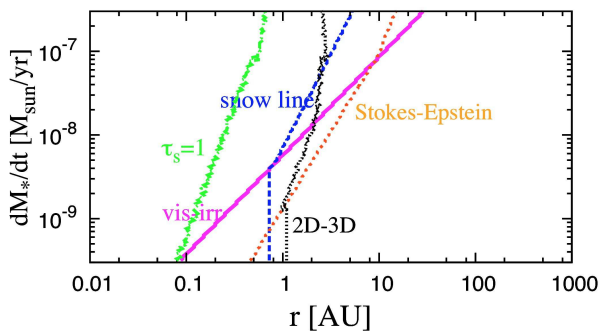}
\caption{Boundaries of disk conditions and pebble accretion modes
in the case of Fig.~\ref{fig:acc01}d.
The Stokes-Epstein (brown dashed line) and 2D-3D (black dotted line)
boundaries are common also in other panels in Fig.~\ref{fig:acc01}.
The boundary of viscously heated and irradiation heated regimes (magenta solid line), 
snow line (blue dashed line),
and $\st = 1$ line (green dashed line) depend on the disk mass accretion rate, $dM_*/dt$.
The left side is characterized by a viscous, 2D accretion, Stokes and high $\st$ regime, while
the right side is characterized by a irradiation, 3D, Epstein and lower $\st$ regime.
}
\label{fig:bd}
\end{figure}

Figure~\ref{fig:acc01} shows $t_{\rm acc}$
calculated by the formulas in section 3.5, 
as a function of $r$ and $\dot{M}_*$ for different disk conditions.
We use $\alpha_3 = 1$ and $\dot{M}_{\rm F4} = \dot{M}_{*8}$.
The Stokes parameter $\st$ is calculated from the prescriptions in section 3.4. 
The planetary embryo mass is set to be $M=0.1\,\mearth$.
In the 3D regime, $t_{\rm acc}$ is independent of $M$, while it increases 
in proportion to $M^{1/3}$ in the 2D case.
In general, the accretion is 2D in the inner disk regions
and 3D in the outer regions (see Fig.~\ref{fig:bd}).
For larger $M$, the 2D region expands toward the outer disk.

Figure~\ref{fig:acc01}a assumes that the entire disk is in the
irradiation regime and $\zeta, \kappa \simeq 1$,
which is an often used setting for pebble accretion calculations. 
The left half ($\la 1\,\au$) is in a 2D-Stokes regime and the other half
is in a 3D-Epstein regime (Fig.~\ref{fig:bd}). 
At $r \la 1 \,\au$, embryos would grow equally in all regions ($\mid \delta \mid \ll 1$),
while they grow in a weak inside-out manner ($\delta \simeq 1$) at $r \ga 1\,\au$.
In Figure~\ref{fig:acc01}b,  
it is assumed that the entire disk is in the viscous regime and $\zeta, \kappa \simeq 1$.
Although the dependence is very weak,
the growth mode is outside-in ($\delta <0$) at $r \la 1\,\au$.
In Figure~\ref{fig:acc01}c, we consider a more realistic model that
includes a transition from an inner viscous regime to an outer irradiation regime as in section 2.
In this case, a fast-growing region is found at $\sim 1\,\au$.
Figures~\ref{fig:acc01}a, b, and c demonstrate how
the disk conditions influence planetary growth by pebble accretion and
ultimately, the configurations of final planetary systems. 

Figure~\ref{fig:acc01}d corresponds to the case
in which the correction factors $\zeta$ (Eq.~(\ref{eq:zeta}))
and $\kappa$ (Eq.~(\ref{eq:kappa})) are included 
along with the viscous and irradiation regimes. 
As shown in section 3.4, since $\st \propto r^{-1-q}$ in the Stokes regime,
$\st$ increases with inward migration even without accounting for a growth of the pebbles.
As shown by Fig.~\ref{fig:bd}, the inner disk regions are
characterized by $\st > 1$ and a 2D-Stokes regime (see Fig.~\ref{fig:bd}).
As a result, $t_{\rm acc}$ is proportional to $\zeta/\kappa$.
While $\kappa^{-1}$ increases because of a decrease of the collision cross section, 
$\zeta$ decreases owing to an increase of pebble surface density caused by
the reduction of $\v_r$,
so that $t_{\rm acc}$ increases or decreases corresponding to
the change in $\zeta$ and $\kappa^{-1}$, compared with
Figure~\ref{fig:acc01}c.
In the limit of large $\st$, the $\kappa^{-1}$-increase dominates over
the $\zeta$-decrease (see the upper left region of Figure~\ref{fig:acc01}d), 
because $\kappa$ is an exponential function.
In order to study the formation of close-in Earths and super-Earths,
a more detailed analysis of pebble accretion rates at $\st \ga 1$ is necessary.

Lastly, Fig.~\ref{fig:acc01}e considers the full model which,
following \citet{Morbidelli15a}, assumes that
the icy mantles of pebbles sublimate inside the snow line and release
mm-sized silicate grains, resulting in a few orders of magnitude reduction of $\st$.
Even if silicate components are present as larger clumps within the icy pebbles, 
collisional fragmentation would decrease their sizes below centimeters
\citep{Birnstiel12, Banzatti15}. 
With the fact that the bouncing barrier would furthermore prevent
their growth beyond millimeter sizes, we set the silicate pebble size to $R=1$mm.
Because of the small $\st$, these small silicate pebbles (or
dust grains) are stirred up and the accretion mode becomes 3D.
As shown in Eq.~(\ref{eq:tacc3D_rdep}), the pebble accretion timescale
in the 3D mode increases as a result of the decrease of $\st$.
The slowing down of the pebble accretion rate inside the snow line
is evident.

As shown in Figure~\ref{fig:bd}, the snow line and transitions from 
viscous to irradiative, Stokes to Epstein and 2D to 3D occur at 
similar orbital distances. 
The location of $\st \sim 1$ is also similar.
The radial dependence of the pebble accretion rate
changes across these boundaries.
Furthermore, since all planetary embryos share the same pebble flux,
if the outer embryos efficiently filter the pebble flux,
the inner embryos cannot grow even if the accretion cross section is larger.
As demonstrated here, configurations of planetary systems formed by pebble accretion 
sensitively depend on the disk conditions.\footnote{Since the Stokes number $\st$ also significantly affects accretion rate,
dust/pebble growth, internal structure, and sublimation are also important factors.}

This means that the outcome of planet formation through pebble
accretion is highly sensitive to the hypotheses made. It also means
that pebble accretion could be responsible for the large diversity of
planetary systems that we observe. 
Differences in disk outer radii,
surface densities and radiative properties would naturally be
generated through the collapse of molecular cloud cores of different
densities and angular momentum \citep[e.g.,][]{HuesoGuillot2005}.
Furthermore, as we discussed in section 3.4, the timescale for 
the pebble formation front to reach the outer edge of the disk (see Eq.~(\ref{eq:tpebdep}))
 depends on disk size and is generally on order of $10^5$\,yr, which is 
much shorter than typical gas disk lifetimes ($\sim$ a few million years).
This implies that the initial total mass of solid materials, $\dot{M}_*$, $\sigg$ and $r_{\rm out}$ 
significantly affects the final configurations of planetary systems.
The formation of planetary systems 
via pebble accretion should depend sensitively on initial disk parameters. 

Detailed descriptions of the disk initial conditions, thermodynamic
properties and evolution are required to correctly predict
and interpret the distributions of exoplanetary systems
by planet population synthesis simulations
\citep[e.g.,][]{IL04, IL08, Mordasini09, IL13, Albert13, Bitsch15b}.
In a separate paper, we show the results of planet population synthesis
simulations based on pebble accretion.

Another important issue is where and how planetary embryos are formed.
Indeed, the formation of seed embryos is essential for an efficient pebble accretion.
Because the $r$-dependence of pebble accretion rate is weak, the initial locations of the embryos
regulate the final planetary systems.
Special sites at which the embryos would form preferentially (e.g.,
the inner edge of the dead zone, the snow line, locations of opacity
jumps) could control the final configurations of planetary systems. 
The initial size of the embryos is also important: For example, 
for embryos smaller than 100 km in radius, pebble accretion is
slower and less efficient than planetesimal accretion (see also Chambers 2016, submitted).

\section{Summary}

We have derived simple analytical formulas for pebble accretion timescales, 
assuming settling regime.
The formulas are explicitly presented in section 3.5.
We next evaluated their radial dependence 
to discuss final configurations of planetary systems formed through
pebble accretion (section 4).

We found that the radial dependence of the pebble accretion rate
is generally relatively weak but that it changes significantly
(including sometimes changing sign) across the boundaries defined by
different regimes, such as the transitions from the viscous-dominated to
the irradiation-dominated disks, from the Epstein drag to the Stokes drag, 
from 2D to 3D accretion regimes, and whether Sthe tokes number 
$\st$ is smaller or larger than $1$.
All of these boundaries, as well as the snow line, occur at
similar distances from the star, $O(1)$\,au.
Inside the snow line,  sublimation of icy mantle of grains 
may change Stokes number by orders of magnitude.
Since the locations of the boundaries depend on
the disk models, we expect intrinsic changes in the properties of the disk to lead to
a large variety of final outcomes.

Self-consistent simulations of planet growth and disk evolution
are much more important for pebble accretion scenario than
for classical planetesimal accretion scenario.
The variety of protoplanetary disks observed combined to
the high sensitivity of pebble accretion on the properties of this
disk indicates that the diversity of planetary systems formed should
be larger than what can be obtained through classical planetesimal accretion.   
This diversity, as predicted by a planet population synthesis model
including pebble accretion, will be discussed in a future paper.

\vspace{1em}
\begin{acknowledgements}
We thank Ramon Brasser and Soko Matsumura for careful proofreading.
We also thank John Chambers for helpful comments as a reviewer.
S. I. is thankful for the hospitality given during his visit to the Observatoire de la C\^ote d'Azur.
This work was partly supported by JSPS KAKENHI Grant \#15H02065.
\end{acknowledgements}

%\begin{thebibliography}{}
\bibliography{pebbles} % your references Yourfile.bib
%\end{thebibliography}

\begin{appendix}

\section{Symbols used in this work}

\begin{table*}[htb]
\caption{Definitions of symbols}\label{tab:symbols}
\begin{tabular}{c | l } \hline
Symbol & Definition \\ \hline \hline
$\st$ & $t_{\rm stop}\Omega$: Stokes number  \\
$\chi $ & $ \sqrt{1+ 4\st^2}/(1+\st^2) $\\
$\zeta$ & $1/(1+\st^2)$ \\
$\kappa$ & cutoff of cross section defined by Eq.~(\ref{eq:kappa}) \\
$\tau_{\rm s, crit1}$ & dust migration is faster than growth for $\st > \tau_{\rm s, crit1}$ (Eq.~(\ref{eq:tau_s_crit1B})) \\
$\tau_{\rm s, crit2}$ & runaway dust growth occurs for $\st > \tau_{\rm s, crit2}$ (Eq.~(\ref{eq:tau_s_crit2})) \\
$r_{\rm snow}$ & snow line at $T\sim 170$K (Eq.~(\ref{eq:r_snow_vis}) or (\ref{eq:r_snow_irr}))\\
$r_{\rm vis-irr}$ & viscously-heated and irradiation boundary (Eq.~(\ref{eq:r_vis_irr}))\\
$r_{\rm ES}$ & Epstein and Stokes drag boundary (Eq.~(\ref{eq:Ep_St}))\\
$r_{\rm pf}$ & pebble formation front \\
$\lambda_{\rm mfp}$ & gas mean free path \\
$\hg$ & $c_s/\Omega$: gas scale height  \\
$\hp$ & $(1 + \st/\alpha)^{-1/2}\hg \sim (\alpha/\st)^{1/2}\hg$: pebble scale height  \\
$\hhg$ & $\hg/r$ \\ 
$\hat{h}_{\rm p}$ & $\hp/r$ \\ 
$\sigg$ & gas surface density  \\
$\Sigma_{\rm p}$ & pebble surface density  \\
$\eta$ & $\v_r - \v_K = (\hhg^2/2) \left| d \ln P/d \ln r \right|$\\
$\eta'$ & $\chi \eta$\\
$\dot{M}$ & pebble mass accretion rate onto a planetary embryo \\
$\dot{M}_{\rm F}$ & pebble mass flux (accretion rate) through the disk \\
$\dot{M}_*$ & gas accretion rate through the disk \\
$\Delta \v$ & relative velocity between the embryo and a pebble \\
$b$ & radius of the collision cross section \\
$M$ & embryo mass \\
$M_{\rm BH}$ & transitional embryo mass between Bondi and Hill regimes (Eq.~\ref{eq:M_BH}) \\
$M_{\rm 2D3D}$ & transitional embryo mass between 2D and 3D accretion  (Eq.~\ref{eq:M2D3D})\\
$R$ & embryo physical radius \\
\hline
$q$ & $ \hhg \propto r^{q}$ \\
        &  \hspace*{1em} $q=1/20$ [viscous],\; $2/7$ [irradiation] \\
$p$ & $\st \propto r^{p}$ \\
$\gamma$ & $T \propto r^{-\gamma}$;  \\
        &  \hspace*{1em} $\gamma = -2q + 1 =9/10$ [viscous],\; 3/7 [irradiation] \\
$\xi$ & $\sigg \propto r^{-\xi}$  \\
        &  \hspace*{1em}  $\xi = 2q + 1/2 =3/5$ [viscous], 15/14 [irradiation] \\
\hline
$ M_{*0}$ & $M_*/1\,\msol$   \\
$ L_{*0}$ & $L_* /1\,\lsol$  \\
$ \alpha_3$ & $\alpha/10^{-3}$  \\
$ \dot{M}_{*8} $ & $\dot{M}_*/10^{-8}\,\msol{\rm yr}^{-1}$  \\
$ \dot{M}_{\rm F4}$ & $\dot{M}_{\rm F}/10^{-4}\,\mearth{\rm yr}^{-1}$  \\

\hline

\end{tabular}
\end{table*}

\end{appendix}

\end{document}